# PhyPro-FM: A Physics and Prompt-Guided Graph Foundation Model for Contingency Analysis

Xu Yang, *Graduate Student Member, IEEE*, Zhanyuan Sheng, Chenhui Lin, *Senior Member, IEEE*
Haotian Liu, *Member, IEEE*, Qi Wang, *Member, IEEE*, Yue Yang, *Member, IEEE*, and Wenchuan Wu, *Fellow, IEEE*

***Abstract*—The continuous evolution of next-generation power systems, characterized by the high penetration of renewable energy resources and dynamic topology variations, poses formidable challenges to traditional N-1/N-2 contingency analysis in terms of computational efficiency and scalability. While deep learning-based methods offer accelerated inference, existing studies predominantly treat contingency analysis as a pure regression problem, often overlooking the physical essence of grid operation and engineering requirements of real-world scenarios. To bridge this gap, this paper proposes a physics and prompt-guided graph foundation model (PhyPro-FM) for generalizable and engineering-viable contingency analysis. The proposed architecture integrates graph neural networks with a learnable global query mechanism and, notably, introduces a prompt-guided learning paradigm. By leveraging historical snapshots as few-shot prompts, the model dynamically captures correlations between target and historical operating conditions, significantly enhancing generalization capabilities under unseen topologies. Furthermore, we design a physics-informed training mechanism that prioritizes branch power flow errors over nodal voltage errors, aligning the optimization objective with critical grid security constraints. To ensure engineering applicability, a physics-informed inference mechanism is also developed, incorporating a residual-based correction module and a specialized post-processing strategy for singular branches to circumvent numerical instability. Comprehensive experiments demonstrate that PhyPro-FM outperforms state-of-the-art methods in accuracy and generalization, validating its superior physical consistency and robustness for practical engineering applications.**

## I. Introduction

With the ongoing evolution of next-generation power systems, an increasing number of distributed energy resources are being integrated into the grid. On the generation side, renewable energy sources represented by wind and solar power are widely connected, gradually replacing traditional large-scale generators [1]-[3]. On the demand side, flexible loads are constantly emerging, such as electric vehicle aggregators, AI servers, and data centers [4]-[6]. Meanwhile, on the grid side, newly built transmission lines, line maintenance, and N-1/N-2 contingencies cause continuous changes in grid topology [7], [8]. The randomness and volatility across these three dimensions are leading to increasingly drastic variations in power flow, simultaneously making situation awareness both more critical and more challenging.

As one of the most fundamental and important tools for power system situation awareness, contingency analysis empowers critical tasks including grid planning, operational mode verification, and economic dispatch [9], [10]. Traditional contingency analysis approaches rely on mathematical methods such as the Newton-Raphson method for iterative solution [11], [12]. Although it can ultimately yield accurate results, it suffers from the drawbacks of long computation time and the inability to perform multi-scenario batch calculations. Under the current power system landscape, real-time, high-precision, and large-scale batch contingency analysis has become imperative, such as interval evaluation of renewable energy output, N-1/N-2 contingency screening, and massive user energy consumption scenarios, posing a formidable challenge to traditional methods.

### A. Literature Review and Gap Analysis

Leveraging their strengths in fast computation and batch inference, deep neural networks (DNNs) have been progressively adopted by the research and engineering community for contingency analysis and related downstream tasks, such as state estimation, N-1/N-2 screening, and optimal power flow [13]. Early efforts constructed multi-layer perceptron (MLP) architectures to enable efficient end-to-end analysis. For example, [14] introduced the application of MLPs to N-1/N-2 screening, validating their substantial potential as a computationally-efficient surrogate for contingency analysis and power security assessment. [15] and [16] extended MLPs to optimal power flow problems, delivering applicable solutions with minimized power loss.

Having established the feasibility of data-driven approaches, the research community has shifted its focus toward integrating physical information into DNN training and inference processes. By embedding physical knowledge through physics-informed neural networks (PINNs), the learning performance and ultimate precision of DNNs can be significantly enhanced. For example, [17]-[19] demonstrated this by applying physical knowledge to contingency analysis, state estimation, and

This work was supported in part by the Beijing Natural Science Foundation under Grant L243003 and the National Science Foundation of China under Grant U24B6009. *(Xu Yang and Zhanyuan Sheng are co-first authors.) (Corresponding author: Wenchuan Wu.)*

X. Yang, Z. Sheng, C. Lin, H. Liu, and W. Wu are with the State Key Laboratory of Power Systems, Department of Electrical Engineering, Tsinghua University, Beijing 100084, China.

Q. Wang is with the Hong Kong Polytechnic University, Hong Kong 999077, China.

Y. Yang is with the State Key Laboratory of High-Efficiency and High-Quality Conversion for Electric Power, Hefei University of Technology, Hefei 230009, China.

optimal power flow, respectively. Other studies, on the other hand, incorporate the grid's graph information as input features into DNNs, thereby extending their capability to perceive and reason about the network topology [20]-[22].

Given the dynamic nature of grid topology and the practical demands of N-1/N-2 screening, merely treating graph information as static features is far from sufficient. Consequently, a growing number of researchers have turned to graph neural networks (GNNs) for topology processing [23]-[29], such as graph convolutional network (GCN) and graph attention network (GAT). By leveraging the message passing mechanism, i.e., iteratively aggregating neighbor features and edge information across nodes, GNNs inherently encode local and even global topological structures into each node's representations. This substantially enhances the model's generalization capability to unseen topologies and its prediction accuracy under dynamic topology variations.

Beyond that, in recent years, graph foundation models have sparked considerable interest due to their powerful representation learning capabilities and few-shot generalization properties [30]. Through self-supervised pre-training on a massive corpus, combined with downstream task fine-tuning, these models can rapidly generalize to unseen systems and diverse downstream application scenarios. This has motivated pioneering efforts to design dedicated architectures and training paradigms. Notably, [31] developed a foundation model using graph transformers (GTs) [32] and self-supervised pre-training, which not only improved the accuracy of contingency analysis but also enabled efficient and fast generalization to downstream tasks and unseen topologies. Microsoft and IBM, on the other hand, proposed GridSFM [33] and GENCO [34], respectively, foundation model architectures designed for contingency analysis and optimal power flow problems, delivering high solution quality and robust cross-scenario generalization on various grid scales and different configurations.

Despite the considerable depth of existing research, these studies still treat contingency analysis and its downstream tasks as pure approximation or regression problems, seeking to achieve better fitting performance through deeper networks, more sophisticated message-passing mechanisms, or larger parameter counts, while overlooking the physical essence of power system operation and engineering requirements in real-world contingency analysis. This oversight of physical essence manifests primarily in the following three perspectives:

1) *The perspective of data utilization mechanism.* The vast majority of existing studies merely treat historical power flow snapshot data as training datasets for models, aiming to approximate the high-dimensional non-linear relationship between nodal power injections and nodal voltages through extensive training. However, the utility and informational value of historical data extend far beyond this narrow scope. On the one hand, analogous to the few-shot learning capability exhibited by general-purpose large language models [35], a contingency foundation model should acquire not only the input-output mapping but also few-shot learning ability, i.e., the capacity to simulate given similar historical snapshots. Furthermore, the power flow residuals and prediction errors of pre-trained models on historical data can also serve as a highly informative source of physical knowledge, helping the model to further refine its coarse predictions.

2) *The perspective of training mechanism.* During training, existing methods excessively pursue the absolute accuracy of voltage magnitudes and phase angles. However, power system operators in practice are more concerned with branch power flows. For contingency analysis in large-scale transmission networks, branch power flow errors are more critical to the core constraints of grid security and stable operation than nodal voltage errors, since branch power flow violations directly lead to line overloading, transformer failures, and even widespread blackouts, whereas small voltage deviations within allowable ranges typically do not trigger immediate severe consequences. Therefore, the training objective should not be solely oriented toward nodal voltage fitting accuracy, but should prioritize branch power flow errors as one important optimizing target, ensuring the physical correctness of power flow distribution during model training.

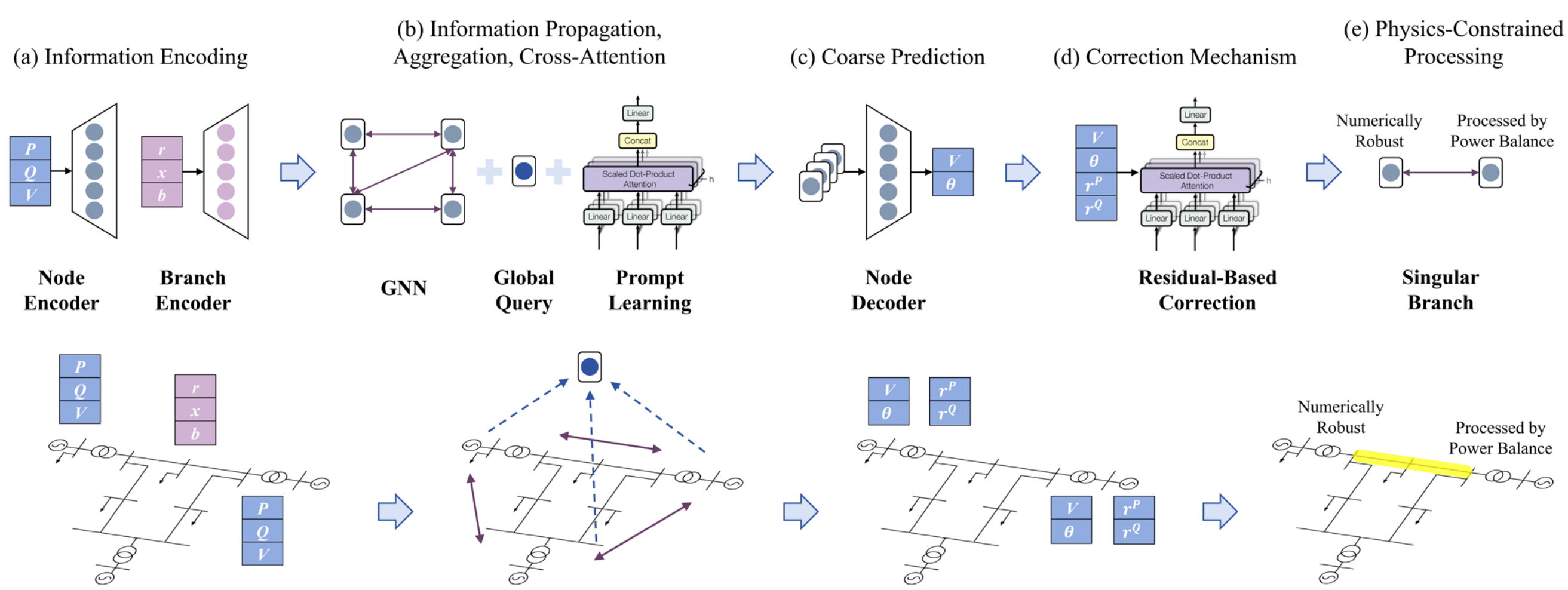

**Fig. 1.** Overall scheme of PhyPro-FM and corresponding methodologies.

3) *The perspective of inference mechanism.* Similar to the previous perspective, most existing works directly treat model inference outputs as final results, assuming that minimized voltage errors will lead to optimized contingency outcomes, without applying any corresponding post-processing methods. Despite the accuracy achieved by machine learning, a significant gap remains in terms of engineering viability, thus calling for tailored post-processing strategies. Also, in real-world power systems, there exists a certain number of "singular branches" with extremely small impedances, where even minor phase angle differences at both terminals can trigger drastic perturbations. Unfortunately, some of these singular branches are non-negligible tie-lines connecting different working areas, and the power flow on them must undergo appropriate assessment; otherwise, the results will be entirely unacceptable and even unusable.

### *B. Contributions*

Therefore, realizing a truly engineering-viable contingency foundation model necessitates innovations and breakthroughs across all three perspectives discussed above, grounded in deep integration of physical essence and real-world engineering requirements. Considering the limitations of existing studies, in this paper, we propose a physics and prompt-guided graph foundation model, PhyPro-FM, for generalizable contingency analysis.

As shown in Fig. 1, the proposed model and corresponding methodologies can be primarily divided into five stages: The first stage involves information encoding, where node encoder and branch encoder project node and branch features into a common latent space. The second stage focuses on information propagation, aggregation, and cross-attention, realized through the stacking of several message passing blocks. This mechanism integrates GNN-based local topology perception with global aggregation driven by a learnable global query. Specifically, it also employs a cross-attention mechanism [36] to align historical snapshots (serving as few-shot prompts) with the current target snapshot. In the third stage, coarse prediction is performed by extracting node embeddings that have undergone information propagation and aggregation, and passing them through a shared prediction decoder to yield initial estimates of nodal voltage magnitudes and phase angles.

The fourth stage then applies a correction mechanism grounded in power flow residuals and historical model errors, in which another cross-attention module receives the coarse predictions along with the associated power flow residuals, while also incorporating the model's historical performance on several past snapshots (also serving as few-shot prompts), thereby further refining the output for current target snapshot. And finally, the fifth stage implements physics-constrained processing for singular branches. For such branches in the network, power flow is not directly derived from terminal voltage magnitudes and phase angle differences as in conventional formulations. Instead, it is determined by enforcing power balance constraints (KCL laws) at the more "numerically robust" of the two terminal nodes, effectively circumventing numerical instability and solution divergence arising from the vanishingly small impedances characteristic of singular branches. Contributions of this paper can be summarized as follows:

1) **PhyPro-FM architecture.** This architecture integrates GNN-based local topology perception with global information aggregation driven by a learnable global query. The key innovation lies in the introduction of a prompt-guided learning mechanism, which leverages historical snapshots as few-shot prompts to guide current predictions. This enables the model to dynamically capture correlations between the target and historical snapshots rather than blindly fitting non-linear mappings. Consequently, it effectively exploits historical physical patterns and significantly enhances generalization capabilities under unseen topologies and operating conditions.

2) **Physics-informed training mechanism.** Departing from traditional methods that focus solely on absolute nodal voltage errors, the proposed physics-informed training mechanism prioritizes branch power flow errors as one important optimization objective to align with the physical essence of grid security. Furthermore, this training strategy is tailored to the proposed PhyPro-FM architecture and prompt-guided learning paradigm, jointly optimizing local and global features to strengthen the model's adherence to critical physical constraints.

3) **Physics-informed inference mechanism.** This mechanism incorporates two post-processing stages: a correction module based on power flow residuals and historical model errors, and a specialized handling strategy for singular branches. This ensures that the model outputs satisfy not only mathematical convergence but also physical laws, thereby avoiding numerical instability and guaranteeing engineering viability in complex scenarios. Comprehensive comparison and ablation studies on different systems demonstrate the effectiveness and superiority of the proposed method.

It should also be emphasized that although the results presented in this paper are based on contingency analysis, the proposed model and the physics and prompt-guided methodology can he readily adapted to multiple downstream tasks, such as optimal power flow and state estimation.

## II. Methods

### *A. Graph Encoding and Prompt Alignment*

To leverage the graph foundation models for batch contingency analysis, the corresponding power grid should first be represented in a graph format $\{\mathcal{N},\mathcal{E},A\}$, where $\mathcal{N}$ is the collection of system nodes; $\mathcal{E}$ is the collection of system branches; and $A$ is the adjacency matrix of the system, capturing the topological connections between nodes. For any node $i \in \mathcal{N}$, its nodal features include: node type (PQ/PV/slack); load $P_{i,d}$ and $Q_{i,d}$; generation $P_{i,g}$ and voltage $V_i$ (for PV nodes); net nodal injection $P_{i,in} = P_{i,g} - P_{i,d}$; nodal shunt admittance; and direct current (DC) power flow prior phase angle $\theta_{i,DC}$. Similarly, for any branch $ij \in \mathcal{E}$, its branch features include: branch type (transformer/transmission line); transformer ratio and direction (for transformer branches); resistance $r_{ij}$; reactance $x_{ij}$; conductance $g_{ij}$; susceptance $b_{ij}$; and line charging susceptance $b_{ij,c}$. To ensure training stability and enhance performance, we employed standard data processing techniques (such as normalization) and feature augmentation strategies (including DC priors and net nodal injections). However, since these pre-processing and augmentation steps are not unique innovations of this work, all methods utilize identical inputs and outputs in the subsequent experiments to ensure a fair comparison.

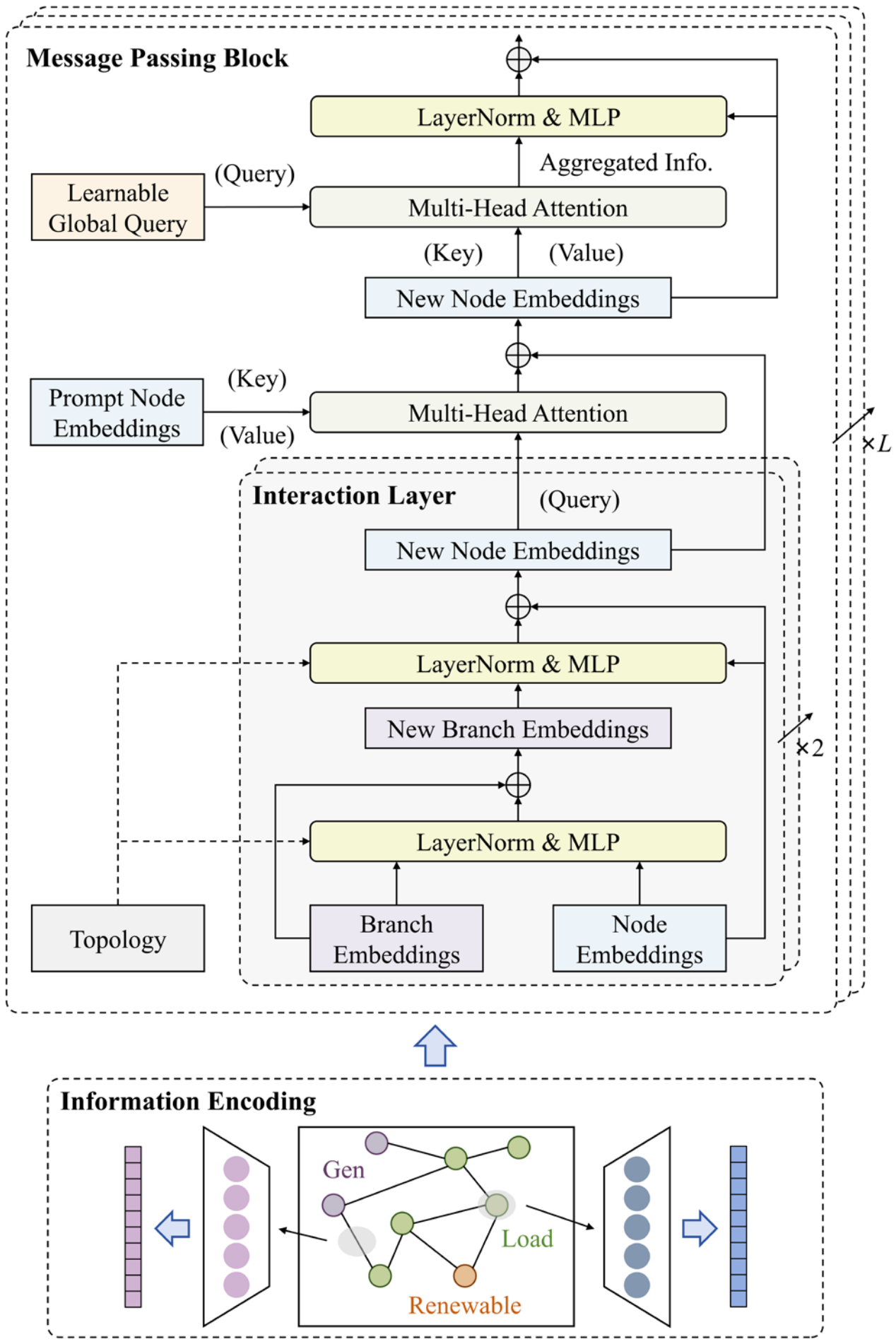

**Fig. 2.** Information processing workflow in PhyPro-FM.

As shown in Fig. 2, the node features and branch features are projected into high-dimensional node embeddings $h_i, \forall i \in \mathcal{N}$ and branch embeddings $e_{ij}, \forall ij \in \mathcal{E}$ by node encoder and branch encoder, respectively. Subsequently, these embeddings, along with the topology relationships, are processed through $L$ message passing blocks to perform information propagation, aggregation, and cross-attention. Each block comprises the following modules:

1) *Interaction layers for local topology perception.*

The initial step involves GNN-based local topology perception, which allows information to propagate along the grid topology, thus capturing topological structures within these embeddings. To better account for branch features, PhyPro-FM deviates from traditional attention-based GNNs and instead constructs interaction layers [37]. Specifically, within the interaction layer, the branch embeddings $e_{ij}$ are first updated by leveraging the node embeddings from its both ends $h_i, h_j$, along with the branch's own embeddings, formulated as:

$$e_{ij}^{(l+1)} = e_{ij}^{(l)} + \text{MLP}\left(\text{LN}\left(e_{ij}^{(l)}, h_i^{(l)}, h_j^{(l)}\right)\right), \forall ij \in \mathcal{E} \quad (1)$$

where LN(·) indicates layer normalization. Afterwards, the node embeddings are also updated using the new embeddings of their connected branches:

$$h_i^{(l+1)} = h_i^{(l)} + \text{MLP}\left(\text{LN}\left(h_i^{(l)}, \sum_{j \in \mathcal{N}(i)} e_{ij}^{(l+1)}\right)\right), \forall i \in \mathcal{N} \quad (2)$$

where $\mathcal{N}(i)$ represents all neighboring nodes of node $i$. Compared to attention-based GNNs, the interaction layer explicitly incorporates branch features into the node embeddings updating process, rather than merely treating them as attention scores, thereby enhancing its graph awareness and expression capabilities.

In PhyPro-FM, to expand the receptive field of nodes, each message passing block contains two consecutive interaction layers. Naturally, the number of layers can be dynamically adjusted according to the grid size.

2) *Prompt-guided historical snapshots alignment.*

Upon completion of local topology perception, PhyPro-FM proceeds to align the current target snapshot with several historical snapshots. This constitutes one of the core innovations of this paper, where historical snapshots from systems with identical topology are introduced not only as training data but also as few-shot prompts to boost the precision of current predictions. In this step, the node embeddings from $k$ historical snapshots along with those from the current target snapshot are jointly fed into a multi-head attention module to construct the prompted information. For node $i$, its prompted information $pro_i$ is calculated as follows:

$$pro_i = \text{MHA}\left(h_i, \{h_i^{t_1}, h_i^{t_2}, \dots, h_i^{t_k}\}, \{h_i^{t_1}, h_i^{t_2}, \dots, h_i^{t_k}\}\right), \forall i \in \mathcal{N} \quad (3)$$

where $\{h_i^{t_1}, h_i^{t_2}, \dots, h_i^{t_k}\}$ are historical snapshots' node embeddings; and $\text{MHA}(Q, K, V)$ is multi-head attention, which aggregates value vectors $V$ using the similarity between query vectors $Q$ and key vectors $K$:

$$\text{Attn} = \text{softmax}\left(\frac{QK^T}{\sqrt{d_K}}\right) \cdot V \quad (4)$$

where $d_k$ is the dimension of these vectors; and softmax(·) is the softmax function. Then the prompted information $pro_i$ is added to the node embeddings $h_i$ to provide more intuition.

3) *Learnable query for global information aggregation.*

Although the aforementioned process effectively integrates local and historical information of the system, contingency analysis is inherently a highly global task. For instance, generation fluctuations from a remote generator can induce changes in global voltage phase angles, while a change in the tap ratio of a tie-line can cause voltage fluctuations across a region. Therefore, it is imperative to integrate and process global information in the contingency foundation model.

In contrast to the burdensome pairwise interactions among nodes in standard GT, we propose a more lightweight method for global information aggregation here. In PhyPro-FM, we construct and maintain a learnable global query $Q^{glo}$ to aggregate global information. After the prompt-guided alignment, the global query, together with the embeddings of all nodes $\{h_i, \forall i \in \mathcal{N}\}$ are also fed into another multi-head attention module to aggregate the global information $glo$:

$$glo = \text{MHA}\left(Q^{glo}, \left\{h_i^{(l)}, \forall i \in \mathcal{N}\right\}, \left\{h_i^{(l)}, \forall i \in \mathcal{N}\right\}\right) \quad (5)$$

Finally, after obtaining the global information, we concatenate it with each node's embeddings and process the results, thereby endowing every node with global awareness:

$$h_i^{(l+1)} = h_i^{(l)} + \mathrm{MLP}\left(\mathrm{LN}\left(h_i^{(l)}, glo\right)\right), \forall i \in \mathcal{N} \quad (6)$$

Different from eq. (3), which aggregates historical information along the temporal dimension, eq. (6) here aggregates information from all nodes along the spatial dimension. Therefore, in eq. (3), node embeddings $h_i$ serve as the query while historical snapshots $\{h_i^{t_1}, h_i^{t_2}, \dots, h_i^{t_k}\}$ act as the key and value. Conversely, in eq. (6), $Q^{glo}$ serves as the query and node embeddings $\{h_i, \forall i \in \mathcal{N}\}$ act as the key and value, realizing a spatiotemporal hybrid attention mechanism.

*B. Model Prediction and Training*

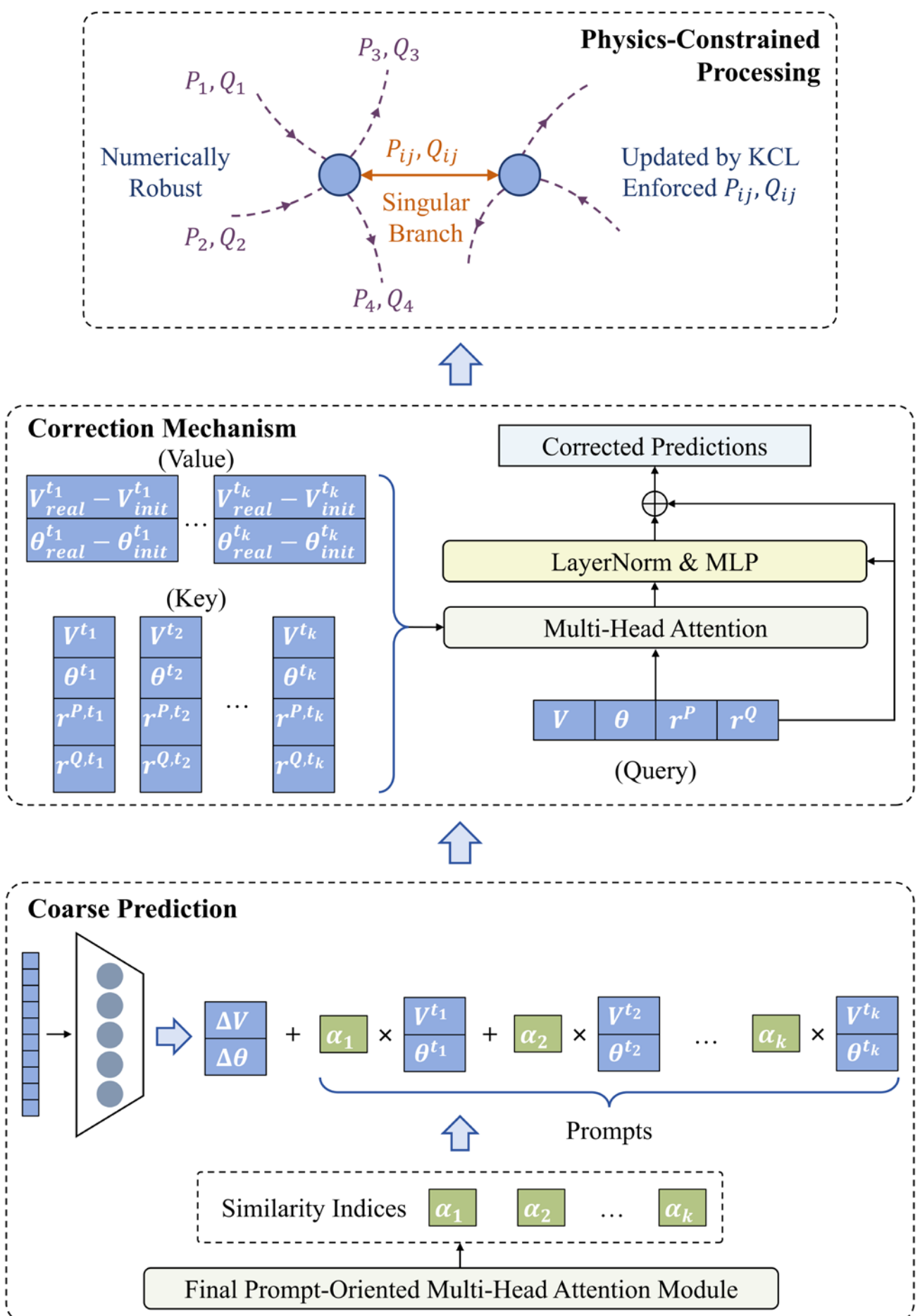

**Fig. 3.** Prediction and post-processing mechanisms.

After information propagation, aggregation, and cross-attention, the resulting node embeddings $\{\tilde{h}_i, \forall i \in \mathcal{N}\}$ are enriched with a comprehensive set of features, including local topology, historical context, and global information. Then the next step is to extract these corresponding embeddings and perform coarse predictions.

As shown in Fig. 3, each node embeddings $\{\tilde{h}_i, \forall i \in \mathcal{N}\}$ are extracted and then passed through a shared prediction decoder to yield the voltage magnitude and phase angle for each respective node $\{(V_i, \theta_i), \forall i \in \mathcal{N}\}$. Conventional approaches typically resort to directly predicting the final values. Nevertheless, such predictions are often blind in nature and can lead to instability issues during the early stage of training. To maximize the utility of prompts, rather than directly forecasting the final values, we guide the model to predict the deviations $\{(\Delta V_i, \Delta\theta_i), \forall i \in \mathcal{N}\}$ between the current target snapshot and historical ones. Specifically, we derive the similarity indices $\{(\alpha_i^1, \alpha_i^2, \dots, \alpha_i^k), \forall i \in \mathcal{N}\}$ between the query vector and key vectors from the final prompt-oriented multi-head attention module within the message passing blocks, which represent the similarity between the current target snapshot and the historical $k$ snapshots. Based on the extracted similarity indices, we obtain a weighted sum of historical power flow results. Thus, the prediction decoder of PhyPro-FM only needs to predict the difference between this weighted sum and the current target snapshot. For node $i$, the initial prediction results of PhyPro-FM $\{(V_{i,init}, \theta_{i,init}), \forall i \in \mathcal{N}\}$ are given by:

$$V_{i,init} = \Delta V_i + \alpha_i^1 V_i^{t_1} + \alpha_i^2 V_i^{t_2} + \dots + \alpha_i^k V_i^{t_k}, \forall i \in \mathcal{N}$$
$$\theta_{i,init} = \Delta\theta_i + \alpha_i^1 \theta_i^{t_1} + \alpha_i^2 \theta_i^{t_2} + \dots + \alpha_i^k \theta_i^{t_k}, \forall i \in \mathcal{N} \quad (7)$$

where $\{(\Delta V_i, \Delta\theta_i), \forall i \in \mathcal{N}\}$ are the outputs from the decoder.

Then the training signals for encoders, message passing blocks, and the prediction decoder are generated by calculating the loss function $\mathcal{L}^{init}$:

$$\mathcal{L}^{init} = \lambda_V \frac{\sum_{i\in\mathcal{N}}\left(V_{i,init}-V_{i,real}\right)^2}{|\mathcal{N}|} + \lambda_\theta \frac{\sum_{i\in\mathcal{N}}\left(\theta_{i,init}-\theta_{i,real}\right)^2}{|\mathcal{N}|} + \lambda_e \frac{\sum_{ij\in\mathcal{E}}\left[\left(\left(P_{ij,init}-P_{ij,real}\right)^2+\left(Q_{ij,init}-Q_{ij,real}\right)^2\right)/|b_{ij}|\right]}{|\mathcal{E}|} + \lambda_{d\theta} \frac{\sum_{ij\in\mathcal{E}}\left(\theta_{ij,init}-\theta_{ij,real}\right)^2}{|\mathcal{E}|} \quad (8)$$

which consists of four components, i.e., voltage magnitude mean squared error (MSE), phase angle MSE, branch power normalized MSE, and branch phase angle difference MSE. Here, $P_{ij,init}$ and $Q_{ij,init}$ denote the active and reactive power flow in branch $ij$ calculated from $\{(V_{i,init}, \theta_{i,init}), \forall i \in \mathcal{N}\}$; $\theta_{ij,init} = \theta_{i,init} - \theta_{j,init}$ is the phase angle difference between node $i$ and $j$; and $\lambda_V, \lambda_\theta, \lambda_e, \lambda_{d\theta}$ are the weight coefficients corresponding to each loss term. The introduction of the latter two physics-informed loss terms effectively embeds physical knowledge into the model's training process, preventing the neglect of physical constraints in existing models, ultimately enhancing the physical consistency and generalization capability of the PhyPro-FM.

*C. Physics-Informed Post-Processing*

The above constitutes the backbone model for contingency analysis. The vast majority of existing models halt at the coarse prediction stage, regarding the direct outputs of the backbone model as the definitive power flow solutions. However, non-negligible power flow residuals and prediction errors persist under specific system states using machine learning algorithms. Therefore, it is necessary to apply physics-informed post-processing to the coarse prediction results.

First, we introduce the residual-based correction. In PhyPro-FM, we construct a correction module that also integrates multi-head attention. Following the coarse prediction, we calculate the power flow residuals for each node on the initial results $\{(r^P_{i,init}, r^Q_{i,init}), \forall i \in \mathcal{N}\}$ and concatenate them as supplementary features to the existing node features. In addition, we calculate the power flow residuals and prediction errors of the backbone model on $k$ historical prompts and provide them to the correction module for reference. The prompts and the current target snapshot are jointly fed into a multi-head attention to analyze their similarity in terms of power flow residuals. Specifically, the current snapshot serves as the query, historical snapshots act as the key, and model errors on historical snapshots act as the value.

Afterwards, the analysis results and the current target snapshot are processed through layer normalization and an MLP to refine the current predictions, yielding the corrected results $\{(V_{i,corr}, \theta_{i,corr}), \forall i \in \mathcal{N}\}$. Then the training signals for the correction module are generated by calculating the loss function $\mathcal{L}^{corr}$:

$$\mathcal{L}^{corr} = \lambda_V \frac{\sum_{i\in\mathcal{N}}(V_{i,corr}-V_{i,real})^2}{|\mathcal{N}|} + \lambda_\theta \frac{\sum_{i\in\mathcal{N}}(\theta_{i,corr}-\theta_{i,real})^2}{|\mathcal{N}|} + \lambda_e \frac{\sum_{ij\in\mathcal{E}}\left[\left[(P_{ij,corr}-P_{ij,real})^2+(Q_{ij,corr}-Q_{ij,real})^2\right]/|b_{ij}|\right]}{|\mathcal{E}|} + \lambda_{d\theta}\frac{\sum_{ij\in\mathcal{E}}(\theta_{ij,corr}-\theta_{ij,real})^2}{|\mathcal{E}|} \quad (9)$$

During training, we adopt a two-stage training framework. In the first stage, we utilize $\mathcal{L}^{init}$ in eq. (8) to train the backbone model, minimizing the error between the initial results and the ground truth. In the second stage, we freeze the backbone model and introduce the correction module, employing $\mathcal{L}^{corr}$ in eq. (9) to train the correction module, further refining the corrected results.

Finally, singular branches may still exist in real-world power systems. Due to their extremely low impedance, even slight errors in phase angle differences can induce significant power flow fluctuations on these branches. Given the inherent precision issues of machine learning methods, relying solely on physics-informed loss and residual-based correction cannot completely eliminate this hidden risk. Moreover, existing studies only focus on the average power flow errors, leaving such singular but critical branches unaddressed. Therefore, dedicated post-processing is required for such branches.

In PhyPro-FM, we propose a physics-constrained post-processing method for singular branches. As shown in Fig. 3, Once a branch $ij \in \mathcal{E}$ is identified as a singular branch requiring post-processing, we calculate sum of the absolute susceptance values $B_i = \sum_{j\in\mathcal{N}(i)}|b_{ij}|$ and $B_j = \sum_{i\in\mathcal{N}(j)}|b_{ji}|$ for its two terminal nodes, indicating their connectivity with other branches and numerical robustness, and a smaller value represents more numerical robust. Afterwards, we select the node with higher robustness as an "anchor node", fixing its voltage magnitude and phase angle $V_{i,corr}, \theta_{i,corr}$. The power flow on branch $ij$ is no longer jointly determined by $V_{i,corr}, \theta_{i,corr}$ and $V_{j,corr}, \theta_{j,corr}$, but is solely governed by the power flow balance constraints (KCL laws) of the selected anchor node. Specifically, the power flow on branch $ij$ equals the total power flowing into the anchor node (such as from other adjacent branches and generator outputs) minus the total power flowing out (such as node loads and shunt admittances). Based on the power flow results of branch $ij$ calculated from KCL laws, we correct the voltage magnitude and phase angle of the non-anchor node. If multiple singular branches are connected, we solve them simultaneously and perform the corrections sequentially. Subsequent experiments demonstrate that the proposed post-processing methods significantly improve the model's accuracy and its performance on branch power flows.

## III. Numerical Studies

### A. Cases and Methods Setup

In this section, in order to verify the effectiveness of the proposed method, we set up 5 open-source systems with different scales, i.e., 39-bus system, 118-bus system, 300-bus system, 1354-bus system, and 3012-bus system as benchmarks. In each case, we randomly perturb system loads, power generation, generator terminal voltages, line parameters, and transformer tap ratios, and generate both normal and N-1/N-2 topologies. Based on this, we comprehensively employ numerical power flow and optimal power flow calculations to construct rich sample data for model training. Training set includes both normal and N-1/N-2 topologies, and a validation set is split from it with a 9:1 ratio. Test set consists of unseen N-1/N-2 topologies. To conserve space, the following subsections only present the results on unseen N-1/N-2 topologies from the test set. The results for validation set (normal topology and seen N-1/N-2) are provided in an online supplementary file [38]. Composition of the training and test sets and hyperparameters of PhyPro-FM in this paper are also provided in the supplementary file [38].

To demonstrate the superiority of the proposed PhyPro-FM, we establish the following comparison baselines: 1) DC power flow; 2) MLP; 3) GCN; 4) GAT; 5) GT; 6) GridSFM, and 7) GENCO. These baselines are trained with the same inputs and outputs as PhyPro-FM, while using voltage magnitude MSE and phase angle MSE as the loss function without physics-informed terms.

In addition, to verify the effectiveness of the proposed method, we also set up three ablation baselines: 1) PhyPro-FM1: We remove all prompt-related components from PhyPro-FM, including prompt-guided learning and correction, to validate the effectiveness of prompts and the necessity of improving historical data utilization efficiency; 2) PhyPro-FM2: We remove the physics-informed terms from PhyPro-FM training, retaining only the voltage magnitude MSE and phase angle MSE, to verify the auxiliary role of physics-informed loss in model training; 3) PhyPro-FM3: We remove all post-processing steps from PhyPro-FM, including residual-based correction and physics-constrained singular branch processing, and directly use the backbone model's output as the final results, to verify the role of post-processing in satisfying physical constraints and ensuring engineering viability.

### B. Intra-System Performance and Comparison

In this subsection, we train and test the proposed PhyPro-FM and baseline methods under two protocols: 1) dedicated training, where the model is trained and tested on the same system; and 2) mixed training, where the model is trained on all 5 systems before testing. The corresponding results, including mean absolute error (MAE) of voltage magnitude, phase angle, branch active power, and maximum absolute error of branch active power, are presented in Table I and Table II, respectively. The best results among all methods are marked in bold.

TABLE I
PERFORMANCE OF PHYPRO-FM AND BASELINE METHODS ON DIFFERENT SYSTEMS (DEDICATED TRAINING)

A. AVERAGE MEAN ABSOLUTE ERROR OF NODE VOLTAGE MAGNITUDE (P.U.)

| Method | 39-Bus | | 118-Bus | | 300-Bus | | 1354-Bus | | 3012-Bus | |
|---|---|---|---|---|---|---|---|---|---|---|
| | N-1 | N-2 | N-1 | N-2 | N-1 | N-2 | N-1 | N-2 | N-1 | N-2 |
| DC | 3.96e-2 | 3.63e-2 | 3.22e-2 | 3.22e-2 | 2.62e-2 | 2.64e-2 | 6.72e-2 | 6.73e-2 | 8.54e-2 | 8.58e-2 |
| MLP | 2.92e-3 | 6.30e-3 | 1.57e-3 | 1.94e-3 | 6.84e-3 | 7.06e-3 | 5.38e-3 | 5.40e-3 | 4.52e-3 | 4.55e-3 |
| GCN | 4.59e-3 | 7.97e-3 | 8.86e-4 | 9.80e-4 | 6.31e-3 | 6.47e-3 | 2.67e-3 | 2.66e-3 | 2.08e-3 | 2.08e-3 |
| GAT | 1.93e-3 | 3.70e-3 | 3.90e-4 | 4.77e-4 | 1.32e-3 | 1.50e-3 | 7.29e-4 | 7.37e-4 | 1.13e-3 | 1.14e-3 |
| GT | 1.50e-3 | 3.14e-3 | 2.68e-4 | 3.68e-4 | 9.20e-4 | 1.09e-3 | 4.54e-4 | 4.63e-4 | 8.54e-4 | 8.66e-4 |
| GridSFM | 1.52e-3 | 3.17e-3 | 1.68e-4 | 2.58e-4 | 4.96e-4 | 6.74e-4 | 1.82e-4 | 2.08e-4 | 2.07e-4 | 2.31e-4 |
| GENCO | 3.51e-4 | **8.58e-4** | 7.61e-5 | 1.04e-4 | 4.90e-4 | 5.79e-4 | 7.28e-4 | 7.38e-4 | 3.36e-3 | 3.33e-3 |
| PhyPro-FM1 | 7.79e-4 | 1.82e-3 | 1.10e-4 | 1.75e-4 | 4.40e-4 | 5.55e-4 | 1.44e-4 | 1.50e-4 | 2.01e-4 | 2.11e-4 |
| PhyPro-FM2 | 4.23e-4 | 9.83e-4 | 7.54e-5 | 9.22e-5 | 3.22e-4 | 3.51e-4 | 1.19e-4 | 1.18e-4 | **1.50e-4** | **1.53e-4** |
| PhyPro-FM3 | 3.99e-4 | 9.50e-4 | 1.12e-4 | 1.33e-4 | 4.08e-4 | 4.42e-4 | 1.28e-4 | 1.28e-4 | 1.77e-4 | 1.81e-4 |
| PhyPro-FM | **3.47e-4** | 8.59e-4 | **7.39e-5** | **9.08e-5** | **3.03e-4** | **3.28e-4** | **1.17e-4** | **1.16e-4** | 1.74e-4 | 1.77e-4 |

B. AVERAGE MEAN ABSOLUTE ERROR OF NODE VOLTAGE PHASE ANGLE (RAD)

| Method | 39-Bus | | 118-Bus | | 300-Bus | | 1354-Bus | | 3012-Bus | |
|---|---|---|---|---|---|---|---|---|---|---|
| | N-1 | N-2 | N-1 | N-2 | N-1 | N-2 | N-1 | N-2 | N-1 | N-2 |
| DC | 1.63e-2 | 2.25e-2 | 4.06e-2 | 4.14e-2 | 0.212 | 0.212 | 4.60e-2 | 4.57e-2 | 4.17e-2 | 4.14e-2 |
| MLP | 3.52e-3 | 6.98e-3 | 1.68e-3 | 2.01e-3 | 4.12e-3 | 4.37e-3 | 1.43e-3 | 1.45e-3 | 1.76e-3 | 1.86e-3 |
| GCN | 4.01e-3 | 5.91e-3 | 2.55e-3 | 3.02e-3 | 7.16e-3 | 7.46e-3 | 2.15e-3 | 2.11e-3 | 3.12e-3 | 3.18e-3 |
| GAT | 3.22e-3 | 4.75e-3 | 1.64e-3 | 2.06e-3 | 5.78e-3 | 5.99e-3 | 1.74e-3 | 1.72e-3 | 2.53e-3 | 2.62e-3 |
| GT | 3.26e-3 | 4.83e-3 | 1.40e-3 | 1.82e-3 | 5.06e-3 | 5.26e-3 | 1.45e-3 | 1.46e-3 | 2.11e-3 | 2.22e-3 |
| GridSFM | 1.27e-3 | 3.19e-3 | 7.71e-4 | 1.18e-3 | 2.05e-3 | 2.58e-3 | 4.71e-4 | 5.45e-4 | 5.15e-4 | 6.14e-4 |
| GENCO | 2.44e-3 | 4.29e-3 | 8.84e-4 | 1.23e-3 | 3.58e-3 | 3.78e-3 | 3.24e-3 | 3.18e-3 | 1.32e-2 | 1.32e-2 |
| PhyPro-FM1 | 1.25e-3 | 2.45e-3 | 5.67e-4 | 9.02e-4 | 1.99e-3 | 2.32e-3 | 4.21e-4 | 4.75e-4 | 5.21e-4 | 5.95e-4 |
| PhyPro-FM2 | 5.80e-4 | 1.15e-3 | 3.96e-4 | 4.74e-4 | 1.02e-3 | 1.07e-3 | **3.39e-4** | **3.35e-4** | **3.76e-4** | **3.95e-4** |
| PhyPro-FM3 | 5.14e-4 | **1.13e-3** | **3.01e-4** | 3.93e-4 | 8.71e-4 | **9.24e-4** | 3.45e-4 | 3.41e-4 | 4.54e-4 | 4.72e-4 |
| PhyPro-FM | **5.05e-4** | **1.13e-3** | **3.01e-4** | **3.87e-4** | **8.61e-4** | 9.26e-4 | 3.44e-4 | 3.38e-4 | 4.58e-4 | 4.76e-4 |

C. AVERAGE MEAN ABSOLUTE ERROR OF BRANCH ACTIVE POWER (P.U.)

| Method | 39-Bus | | 118-Bus | | 300-Bus | | 1354-Bus | | 3012-Bus | |
|---|---|---|---|---|---|---|---|---|---|---|
| | N-1 | N-2 | N-1 | N-2 | N-1 | N-2 | N-1 | N-2 | N-1 | N-2 |
| DC | 0.121 | 0.133 | 0.128 | 0.128 | 0.177 | 0.177 | 1.37 | 1.37 | 0.778 | 0.782 |
| MLP | 6.91e-2 | 0.111 | 7.92e-3 | 9.24e-3 | 3.97e-2 | 4.12e-2 | 8.43e-2 | 8.54e-2 | 4.13e-2 | 4.13e-2 |
| GCN | 0.130 | 0.188 | 1.93e-2 | 2.18e-2 | 8.56e-2 | 8.83e-2 | 0.258 | 0.253 | 0.422 | 0.418 |
| GAT | 9.37e-2 | 0.132 | 1.22e-2 | 1.39e-2 | 6.33e-2 | 6.61e-2 | 0.251 | 0.247 | 0.398 | 0.396 |
| GT | 9.66e-2 | 0.141 | 1.18e-2 | 1.42e-2 | 7.07e-2 | 7.63e-2 | 0.222 | 0.219 | 0.324 | 0.324 |
| GridSFM | 4.88e-2 | 8.60e-2 | 7.13e-3 | 8.80e-3 | 2.70e-2 | 3.17e-2 | 7.74e-2 | 8.01e-2 | 0.178 | 0.184 |
| GENCO | 4.80e-2 | 6.85e-2 | 6.14e-3 | 7.10e-3 | 6.19e-2 | 6.37e-2 | 1.03 | 1.03 | 0.852 | 0.850 |
| PhyPro-FM1 | 1.22e-2 | 2.27e-2 | 1.11e-3 | 1.63e-3 | 2.76e-3 | 3.51e-3 | **3.23e-3** | **3.36e-3** | **2.46e-3** | **2.52e-3** |
| PhyPro-FM2 | 1.39e-2 | 2.26e-2 | 2.32e-3 | 2.50e-3 | 6.19e-3 | 6.36e-3 | 1.76e-2 | 1.73e-2 | 1.51e-2 | 1.51e-2 |
| PhyPro-FM3 | 6.81e-3 | 1.23e-2 | 1.10e-3 | 1.27e-3 | 2.97e-3 | 3.14e-3 | 5.91e-3 | 5.85e-3 | 4.00e-3 | 4.01e-3 |
| PhyPro-FM | **5.18e-3** | **9.68e-3** | **7.72e-4** | **9.23e-4** | **1.79e-3** | **1.92e-3** | 3.70e-3 | 3.66e-3 | 2.62e-3 | 2.62e-3 |

D. AVERAGE MAXIMUM ABSOLUTE ERROR OF BRANCH ACTIVE POWER (P.U.)

| Method | 39-Bus | | 118-Bus | | 300-Bus | | 1354-Bus | | 3012-Bus | |
|---|---|---|---|---|---|---|---|---|---|---|
| | N-1 | N-2 | N-1 | N-2 | N-1 | N-2 | N-1 | N-2 | N-1 | N-2 |
| DC | 0.635 | 0.668 | 1.69 | 1.69 | 13.4 | 13.2 | 1.07e2 | 1.07e2 | 2.12e2 | 2.13e2 |
| MLP | 0.455 | 0.686 | 9.72e-2 | 0.119 | 3.64 | 3.79 | 8.78 | 8.95 | 12.8 | 12.7 |
| GCN | 0.704 | 0.947 | 0.372 | 0.419 | 5.13 | 5.31 | 16.4 | 15.8 | 63.6 | 61.1 |
| GAT | 0.547 | 0.693 | 0.298 | 0.344 | 2.97 | 3.04 | 17.5 | 16.7 | 54.2 | 53.4 |
| GT | 0.602 | 0.800 | 0.333 | 0.402 | 3.57 | 4.09 | 18.5 | 17.3 | 35.5 | 34.3 |
| GridSFM | 0.439 | 0.655 | 0.166 | 0.199 | 1.68 | 1.87 | 3.37 | 3.80 | 22.7 | 24.8 |
| GENCO | 0.286 | 0.339 | 0.107 | 0.122 | 3.12 | 3.12 | 59.2 | 57.7 | 41.4 | 41.5 |
| PhyPro-FM1 | 0.101 | 0.152 | 1.36e-2 | 1.96e-2 | 5.44e-2 | 7.61e-2 | **5.62e-2** | 6.84e-2 | 5.99e-2 | 7.16e-2 |
| PhyPro-FM2 | 6.64e-2 | 0.103 | 2.90e-2 | 2.91e-2 | 8.39e-2 | 8.46e-2 | 0.428 | 0.415 | 0.476 | 0.477 |
| PhyPro-FM3 | 4.35e-2 | 6.81e-2 | 1.26e-2 | 1.49e-2 | 0.104 | 0.108 | 0.157 | 0.160 | 0.276 | 0.265 |
| PhyPro-FM | **2.90e-2** | **4.38e-2** | **8.15e-3** | **9.95e-3** | **2.39e-2** | **2.62e-2** | 5.71e-2 | **5.85e-2** | **5.79e-2** | **5.85e-2** |

First, we examine the voltage magnitude performance of each method. We categorize all methods into three groups: traditional machine learning models (MLP, GCN, and GAT), existing graph foundation models (GT, GridSFM, and GENCO), and the proposed PhyPro-FM along with its ablation variants. As shown in Table I-A, traditional machine learning methods achieve a voltage magnitude MAE on the order of $10^{-3}$ across all cases, whereas graph foundation models perform at the $10^{-4}$ level. Proposed PhyPro-FM and its ablation variants yield the best results in nearly all cases. Notably, PhyPro-FM achieves best performance of $7.39\times10^{-5}$ in the unseen N-1 scenarios of the 118-bus system, while its worst performance of $8.59\times10^{-4}$ occurs in the unseen N-2 scenarios of the 39-bus system. This is attributed to the more severe voltage fluctuations caused by N-2 contingencies in smaller systems. Although the above observations validate the superiority of graph foundation models and the proposed method over traditional machine learning, no strong conclusion can be drawn that graph foundation models hold an absolute advantage in practical applications, since a voltage magnitude accuracy in the range of $10^{-3}$ to $10^{-4}$ is already sufficient for engineering purposes.

Second, we analyze the phase angle performance. As can be seen in Table I-B, graph foundation models perform slightly better than traditional machine learning models in terms of phase angle, except for GENCO, which exhibits a significant degradation on the 3012-bus systems. Nevertheless, PhyPro-FM and its ablation variants remain optimal across all cases, validating its robustness and effectiveness.

Another interesting observation is that PhyPro-FM2 sometimes outperforms PhyPro-FM in terms of voltage magnitude and phase angle. This is because it excludes physics-informed loss terms, allowing the model to focus exclusively on fitting voltage and phase angle. However, the subsequent comparison and ablation studies on branch power flow will demonstrate the necessity of these physics-informed loss terms.

Next, we analyze the branch power flow results. As previously mentioned, accurate fitting of voltage magnitude and phase angle does not necessarily guarantee accurate branch power flow estimation. This is also validated by the performance of PhyPro-FM2. Like all comparison baselines, it employs only MSE loss for voltage magnitude and phase angle. Although its results still outperform all comparison baselines across all cases (demonstrating the effectiveness of the proposed architecture and prompt-guided learning from another perspective), its branch power flow error is nearly an order of magnitude higher than that of the complete PhyPro-FM in some cases. Regarding the maximum branch power error, several comparison baselines exhibit errors as high as tens of p.u. Such large deviations would render the results entirely unacceptable, especially if they occur on critical tie-lines.

Post-processing, including residual-based correction and physics-constrained singular branch processing, also plays a significant role in reducing branch power errors. By comparing PhyPro-FM with PhyPro-FM3, it can be observed that the proposed post-processing methods reduce the branch power MAE by 20%-40% and the maximum absolute error by 35%-80%, thereby improving the physics satisfaction rate.

The results of mixed training are similar to those of dedicated training. PhyPro-FM and its ablation variants achieve the best performance in terms of voltage magnitude and phase angle across all cases, validating the effectiveness of the proposed method. Meanwhile, PhyPro-FM itself attains the best results in branch power across all cases, which further verifies the performance gains contributed by each ablation component.

TABLE II

PERFORMANCE OF PHYPRO-FM AND BASELINE METHODS ON DIFFERENT SYSTEMS (MIXED TRAINING)

A. AVERAGE MEAN ABSOLUTE ERROR OF NODE VOLTAGE MAGNITUDE (P.U.)

| Method | 39-Bus | | 118-Bus | | 300-Bus | | 1354-Bus | | 3012-Bus | |
|---|---|---|---|---|---|---|---|---|---|---|
| | N-1 | N-2 | N-1 | N-2 | N-1 | N-2 | N-1 | N-2 | N-1 | N-2 |
| GCN | 4.84e-3 | 8.13e-3 | 1.05e-3 | 1.16e-3 | 6.34e-3 | 6.50e-3 | 2.85e-3 | 2.83e-3 | 2.37e-3 | 2.37e-3 |
| GAT | 1.77e-3 | 3.45e-3 | 4.93e-4 | 5.93e-4 | 1.32e-3 | 1.50e-3 | 8.14e-4 | 8.13e-4 | 1.34e-3 | 1.35e-3 |
| GT | 1.56e-3 | 3.43e-3 | 3.38e-4 | 4.58e-4 | 9.84e-4 | 1.16e-3 | 5.89e-4 | 5.99e-4 | 1.03e-3 | 1.04e-3 |
| GridSFM | 1.26e-3 | 2.58e-3 | 2.22e-4 | 3.36e-4 | 6.10e-4 | 7.74e-4 | 2.80e-4 | 3.03e-4 | 3.38e-4 | 3.65e-4 |
| GENCO | 1.24e-3 | 2.19e-3 | 4.41e-4 | 4.92e-4 | 2.01e-3 | 2.10e-3 | 2.02e-3 | 2.01e-3 | 4.90e-3 | 4.86e-3 |
| PhyPro-FM1 | 8.89e-4 | 1.81e-3 | 1.60e-4 | 2.15e-4 | 5.74e-4 | 6.78e-4 | 2.22e-4 | 2.25e-4 | 3.81e-4 | 3.93e-4 |
| PhyPro-FM2 | 3.24e-4 | 6.68e-4 | **8.66e-5** | **9.95e-5** | **3.17e-4** | **3.33e-4** | 1.47e-4 | 1.46e-4 | **2.37e-4** | **2.40e-4** |
| PhyPro-FM3 | 3.07e-4 | 6.58e-4 | 1.32e-4 | 1.47e-4 | 4.40e-4 | 4.61e-4 | 1.40e-4 | 1.41e-4 | 2.59e-4 | 2.63e-4 |
| PhyPro-FM | **2.52e-4** | **5.51e-4** | 9.12e-5 | 1.04e-4 | 3.49e-4 | 3.68e-4 | **1.30e-4** | **1.30e-4** | 2.56e-4 | 2.59e-4 |

B. AVERAGE MEAN ABSOLUTE ERROR OF NODE VOLTAGE PHASE ANGLE (RAD)

| Method | 39-Bus | | 118-Bus | | 300-Bus | | 1354-Bus | | 3012-Bus | |
|---|---|---|---|---|---|---|---|---|---|---|
| | N-1 | N-2 | N-1 | N-2 | N-1 | N-2 | N-1 | N-2 | N-1 | N-2 |
| GCN | 4.08e-3 | 5.84e-3 | 3.18e-3 | 3.73e-3 | 7.67e-3 | 8.08e-3 | 2.93e-3 | 2.85e-3 | 3.93e-3 | 3.97e-3 |
| GAT | 3.47e-3 | 5.26e-3 | 2.01e-3 | 2.43e-3 | 6.20e-3 | 6.49e-3 | 2.40e-3 | 2.37e-3 | 3.05e-3 | 3.12e-3 |
| GT | 3.36e-3 | 4.90e-3 | 1.80e-3 | 2.24e-3 | 5.49e-3 | 5.70e-3 | 1.91e-3 | 1.91e-3 | 2.53e-3 | 2.62e-3 |
| GridSFM | 1.26e-3 | 2.67e-3 | 8.64e-4 | 1.27e-3 | 2.85e-3 | 3.43e-3 | 7.39e-4 | 7.94e-4 | 7.52e-4 | 8.37e-4 |
| GENCO | 4.74e-3 | 6.11e-3 | 3.93e-3 | 4.38e-3 | 1.27e-2 | 1.30e-2 | 7.29e-3 | 7.16e-3 | 1.18e-2 | 1.17e-2 |
| PhyPro-FM1 | 1.32e-3 | 2.34e-3 | 9.10e-4 | 1.22e-3 | 2.44e-3 | 2.74e-3 | 7.08e-4 | 7.68e-4 | 8.75e-4 | 9.54e-4 |
| PhyPro-FM2 | 5.42e-4 | 1.03e-3 | 4.65e-4 | 5.48e-4 | **1.04e-3** | 1.08e-3 | 4.69e-4 | 4.59e-4 | **5.38e-4** | **5.55e-4** |
| PhyPro-FM3 | 4.97e-4 | 8.80e-4 | **4.14e-4** | 4.78e-4 | **1.04e-3** | **1.06e-3** | **4.35e-4** | **4.26e-4** | 6.25e-4 | 6.37e-4 |
| PhyPro-FM | **4.89e-4** | **8.37e-4** | 4.20e-4 | **4.77e-4** | **1.04e-3** | **1.06e-3** | 4.38e-4 | 4.28e-4 | 6.33e-4 | 6.46e-4 |

C. AVERAGE MEAN ABSOLUTE ERROR OF BRANCH ACTIVE POWER (P.U.)

| Method | 39-Bus | | 118-Bus | | 300-Bus | | 1354-Bus | | 3012-Bus | |
|---|---|---|---|---|---|---|---|---|---|---|
| | N-1 | N-2 | N-1 | N-2 | N-1 | N-2 | N-1 | N-2 | N-1 | N-2 |
| GCN | 0.134 | 0.195 | 2.31e-2 | 2.60e-2 | 9.48e-2 | 9.71e-2 | 0.364 | 0.358 | 0.680 | 0.678 |
| GAT | 9.73e-2 | 0.141 | 1.55e-2 | 1.77e-2 | 7.62e-2 | 7.95e-2 | 0.362 | 0.356 | 0.552 | 0.548 |
| GT | 0.110 | 0.153 | 1.54e-2 | 1.77e-2 | 8.11e-2 | 8.75e-2 | 0.291 | 0.288 | 0.422 | 0.420 |
| GridSFM | 5.58e-2 | 0.101 | 7.41e-3 | 9.28e-3 | 3.51e-2 | 4.00e-2 | 0.114 | 0.118 | 0.243 | 0.248 |
| GENCO | 0.167 | 0.208 | 4.49e-2 | 4.67e-2 | 0.274 | 0.280 | 1.95 | 1.95 | 1.53 | 1.53 |
| PhyPro-FM1 | 1.36e-2 | 2.51e-2 | 1.66e-3 | 2.16e-3 | 3.49e-3 | 4.15e-3 | 5.19e-3 | 5.26e-3 | 4.31e-3 | 4.35e-3 |
| PhyPro-FM2 | 1.19e-2 | 1.82e-2 | 2.58e-3 | 2.76e-3 | 6.62e-3 | 6.68e-3 | 2.60e-2 | 2.57e-2 | 2.35e-2 | 2.34e-2 |
| PhyPro-FM3 | 5.97e-3 | 1.01e-2 | 1.40e-3 | 1.54e-3 | 3.47e-3 | 3.63e-3 | 6.71e-3 | 6.62e-3 | 5.67e-3 | 5.64e-3 |
| PhyPro-FM | **4.74e-3** | **8.47e-3** | **1.03e-3** | **1.16e-3** | **2.17e-3** | **2.30e-3** | **4.62e-3** | **4.58e-3** | **3.61e-3** | **3.60e-3** |

D. AVERAGE MAXIMUM ABSOLUTE ERROR OF BRANCH ACTIVE POWER (P.U.)

| Method | 39-Bus | | 118-Bus | | 300-Bus | | 1354-Bus | | 3012-Bus | |
|---|---|---|---|---|---|---|---|---|---|---|
| | N-1 | N-2 | N-1 | N-2 | N-1 | N-2 | N-1 | N-2 | N-1 | N-2 |
| GCN | 0.786 | 1.08 | 0.472 | 0.557 | 4.76 | 4.97 | 22.0 | 20.8 | 93.3 | 91.8 |
| GAT | 0.600 | 0.883 | 0.338 | 0.402 | 3.68 | 4.02 | 25.1 | 24.0 | 68.8 | 67.7 |
| GT | 0.649 | 0.875 | 0.380 | 0.444 | 5.04 | 5.76 | 20.8 | 20.4 | 47.5 | 45.2 |
| GridSFM | 0.443 | 0.649 | 0.152 | 0.204 | 2.51 | 2.92 | 4.57 | 5.62 | 27.5 | 30.3 |
| GENCO | 0.822 | 1.05 | 0.653 | 0.668 | 20.2 | 20.8 | 93.5 | 93.7 | 98.4 | 97.8 |
| PhyPro-FM1 | 0.138 | 0.231 | 1.79e-2 | 2.29e-2 | 5.88e-2 | 7.55e-2 | 9.45e-2 | 0.102 | 9.43e-2 | 0.101 |
| PhyPro-FM2 | 5.94e-2 | 9.21e-2 | 3.01e-2 | 3.16e-2 | 9.06e-2 | 9.15e-2 | 0.716 | 0.710 | 0.690 | 0.691 |
| PhyPro-FM3 | 3.75e-2 | 5.91e-2 | 1.54e-2 | 1.67e-2 | 0.113 | 0.114 | 0.150 | 0.151 | 0.374 | 0.378 |
| PhyPro-FM | **3.24e-2** | **5.44e-2** | **1.02e-2** | **1.16e-2** | **2.78e-2** | **3.07e-2** | **7.80e-2** | **7.81e-2** | **7.38e-2** | **7.40e-2** |

A notable observation arises from comparing dedicated and mixed training regimes. PhyPro-FM exhibits a marginal performance drop on the 118-bus to 3012-bus systems but achieves a slight gain on the 39-bus system through mixed training. On the one hand, this highlights the generalization capability of the proposed method, enabling simultaneous training and execution across a wide range of systems. On the other hand, the discrepancy among cases may be attributed to the limited number of training samples in the 39-bus system, where dedicated training has not yet converged to the optimal state, thus requiring supplementary general knowledge from other systems. The subsequent fine-tuning analysis will also validate the effectiveness of this extrapolation capability.

### C. Extrapolation Capability and Fine-Tuning Analysis

In this subsection, we analyze the extrapolation capability of the models. Each method is pre-trained on 39-bus, 118-bus, 300-bus, and 3012-bus systems, and then fine-tuned and tested on the 1354-bus system. The results are shown in Fig. 4- Fig. 5.

As can be observed, PhyPro-FM exhibits superior performance over the comparison baselines, even showcasing a certain degree of zero-shot generalization. This advantage stems from our prompt-guided learning paradigm, which effectively mitigates the zero-shot challenge by framing it as a few-shot task to some extent. This is a pragmatic assumption, as any power grid, regardless of its data scarcity for pre-training, will definitely have operational records available as prompts. This finding further underscores the importance of investigating how to effectively leverage historical or fine-tuning data. Additionally, post-processing techniques also aid in enhancing the overall accuracy. As the number of fine-tuning samples increases, the performance of all models improves steadily. After approximately 1,000 training samples, the voltage, phase angle, and branch power flow errors of PhyPro-FM reach a level that is acceptable for engineering applications.

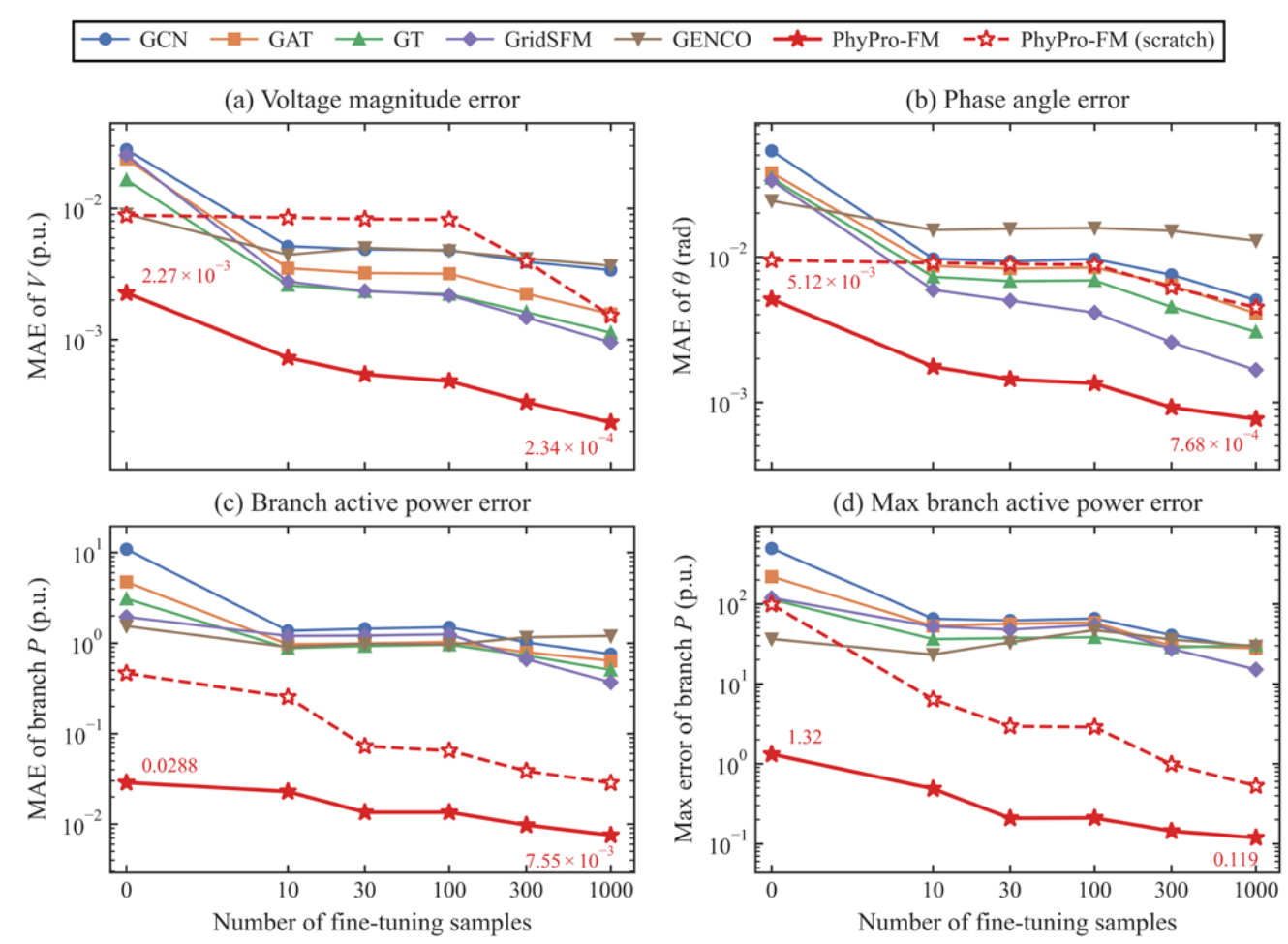

**Fig. 4.** Fine-tuning performance of different methods on unseen N-1 scenarios in 1354-bus system.

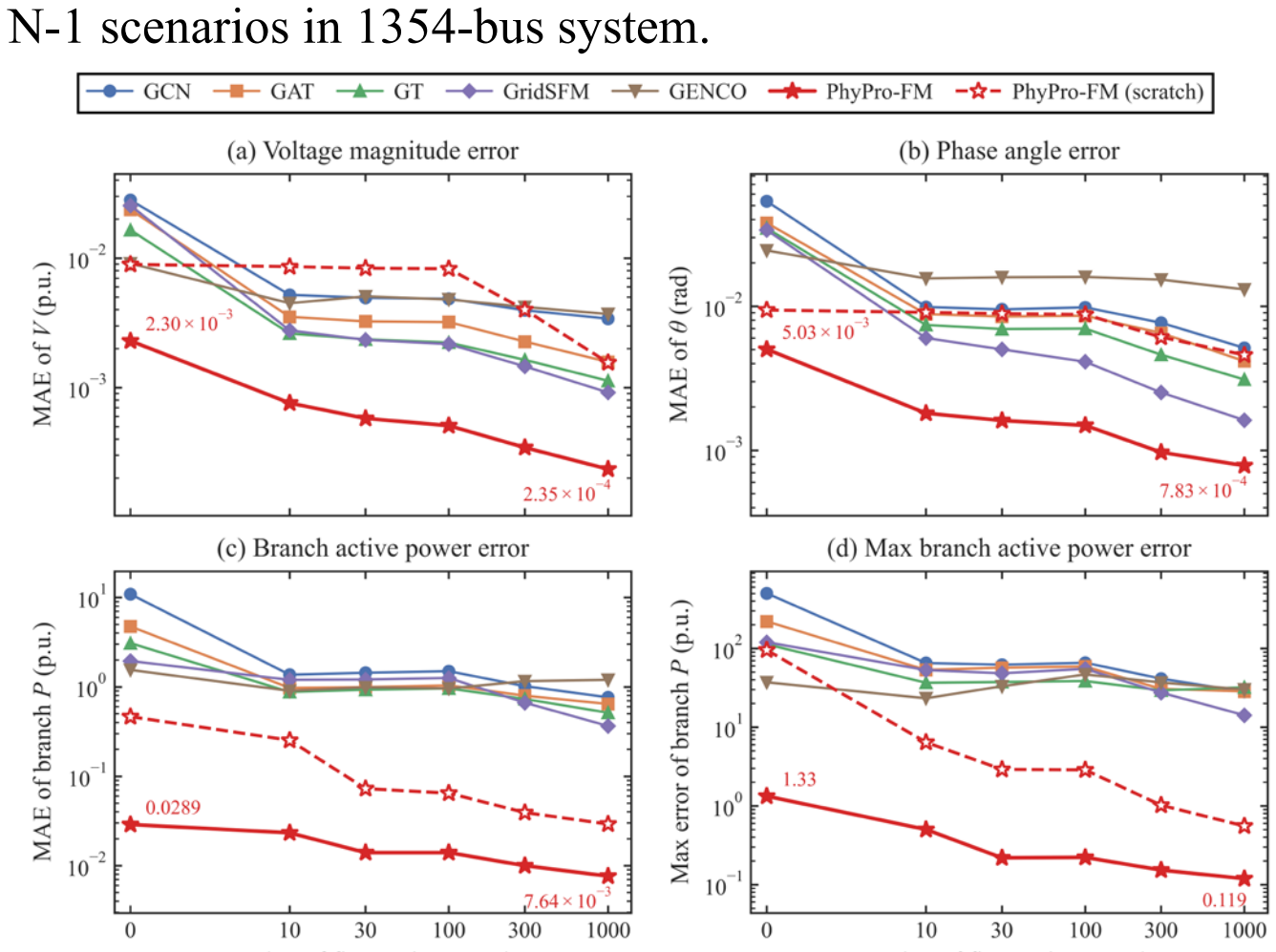

**Fig. 5.** Fine-tuning performance of different methods on unseen N-2 scenarios in 1354-bus system.

On the other hand, we compare the pre-trained PhyPro-FM with a model trained from scratch, denoted as PhyPro-FM (scratch). The results demonstrate that pre-training effectively enhances the model's performance and adaptability, which is consistent with previous research findings.

## IV. Conclusion

In this paper, we develop a contingency foundation PhyPro-FM, which effectively incorporates local topology, global information, historical snapshots, and physical constraints, achieving high accuracy and strong generalization capability.

The results of this paper demonstrate that, beyond pursuing more complex models, delving into the physical essence and engineering requirements of power systems is even more critical. Building upon the proposed physics and prompt-guided framework, exploring additional applications represents a highly important and promising direction for future research.